# The Australian Phillips curve and more

Ivan Kitov, Oleg Kitov

Abstract
A quantitative model is presented linking the rate of inflation and unemployment to the change in the level of labor force. The link between the involved variables is a linear one with all coefficients of individual and generalized models obtained empirically. To achieve the best fit between measured and predicted time series cumulative curves are used as a simplified version of the 1-D boundary elements method. All models for Australia are similar to those obtained for the US, France, Japan and other developed countries and thus validate the concept and related quantitative model.

## Introduction

Five years ago we introduced a concept linking by linear and lagged relationships price inflation and unemployment in developed countries to the change rate of labor force (Kitov, 2006). Corresponding model is a completely deterministic one with the change in labor force being the only driving force causing all variations in the pair unemployment/inflation. Since 2006, many empirically estimated models were (conditional on the length of relevant time series) tested econometrically and demonstrated the presence of cointegrating relations.

Our model is a natural extension of the conventional economic/econometric models and concepts. For example, Stock and Watson (2008) conducted a formal study of a large number of economic/financial variables and indices as predictors of inflation using the Phillips curve. The change in labor force was not included in this enormously large set of approximately 200 predictors. Therefore, we extended this studied set and conducted a similar statistical investigation. Fortunately, this measurable macroeconomic variable is characterized by a much better predictive power and inflation is now not "hard to forecast". The change in labor force in the biggest developed countries (the USA, Japan, France, Germany, etc.) is so good a predictor that there is no need to use any autoregressive (AR) properties of inflation. In this sense, our model is fully deterministic and the model residuals are related to measurement errors.

In its original form, the model was revealed and formulated for the United States (Kitov, 2006). The root-mean-square forecasting error (RMSFE) of inflation at a 2.5 year horizon was of 0.8% between 1965 and 2004. Well-known non-stationary behavior of all involved variables required testing for the presence of cointegrating relations (Kitov, Kitov, and Dolinskaya; 2007).

Both, the Engle-Granger and Johansen approaches have shown the existence of cointegration between unemployment, inflation and the change in labor force, i.e. the presence of long-term equilibrium (in other words, deterministic or causal) relations. Because the change in labor force is likely a stochastic process and it drives the other two variables they also can be formally stochastic process, but fully deterministic ones.

In this paper, we empirically estimate several models of inflation and unemployment for Australia. This is one of few countries we have skipped in our previous analysis and it is good time to fill this gap. Our research and its results add to the extensive literature related to inflation and unemployment in Australia. For example, Norman and Richards (2010) reviewed a range of single-equation models of inflation for Australia. They found that the expectations-augmented standard Phillips curve or mark-up models show better predictive power than the New-Keynesian Phillips curve. Another important conclusion is that the unemployment rate better fits inflation data than either the output gap or level of real marginal costs. Due to the overall decrease in the rate of inflation the relative explanatory power of all inflation models has fallen. Not surprisingly, such determinants of inflation as commodity prices or growth rate of money do not influence Australian inflation. We presume that the absence of labor force in the set of predictors is the cause of the eternal problems in prediction.

Econometrically, price inflation is often considered as a stochastic process. Stock and Watson (2008, 2010) carried out a detailed study of AR properties of many inflation time series in the U.S. Karunaratne and Bhar (2010) used Markov regime switching heteroscedasticity model to capture long-run stochastic trend and short-run noisy components in the rate of Australian price inflation and inflation uncertainty during the post-float era 1983Q3-2006Q4. They found a significant deviation from the mainstream Friedman paradigm on inflation and its uncertainty. However, they linked the inflation dynamics to the macroeconomic policies pursued to achieve external and internal balance as implied by Keynesian Mundell-Fleming model. Our findings partially support this conclusion but the only break in the inflation dynamics has been classified as the change in monetary policy of the Reserve Bank of Australia around 1993 to explicit inflation targeting (Bernanke et al, 1999; Grenville, 1997).

For an open economy, the influence of external factors is an important problem. Do we really see any external shocks in the domestic rate of inflation in Australia? This question was addressed by Buncic and Malecky (2007). They found that domestic demand and supply shocks affect the rate of inflation the most. It was also found that domestic inflation affected by exchange rate shocks. We share this conclusion to the extent all these processes can influence the level of labor force.

Naturally, a sound and quantitatively accurate model of price inflation allows forecasting at various time horizons. Our deterministic model is able to use labor force projections as a predictor, with more accurate projections providing better forecasts. Other inflation models use various parameters. For example, Robinson *et al.* (2003) considered the Phillips curve based on estimates of the output gap and found only poor predictive power relative even to simplest autoregressive model and random walk models (Atkeson and Ohanian, 2001). Strangely, they concluded that the output-gap-based Phillips curves "may continue to be useful in real time". Hall and Jaaskela (2009) found a measurable difference in inflation volatility and thus in the accuracy of prediction between countries with and without inflation targeting. This may be a manifestation of the interaction between inflation and unemployment under our general framework linking both variables to the change in labor force. This effect was clearly observed in France. A much lower rate of price inflation is achieved at the expense of very high rate of unemployment (Kitov, 2007).

The rate of unemployment is a crucial macroeconomic variable as well. The dynamics of unemployment in Australia, as in all developed countries, has no empirically accurate explanation in the mainstream economics and is chiefly based on pure theoretical assumptions on labor market. Karanassou and Sala (2009) presented a multi-equation labor market model comprising labor demand, wage setting and labor supply equations. The main goal was to describe the most prominent changes in the rate of unemployment: the high-amplitude increase between 1973 and 1983 and the fall from 1993 to 2006. Wage, oil price shocks and interest rate were found to cause the rise in unemployment in the 1970s and early 1980s. On the contrary, the fall in the 1990s was explained as driven by the acceleration in capital accumulation. Bardsen *et al.* (2010) also modeled the rate of unemployment in Australia as an asymmetric and non-linear function of aggregate demand, productivity, real wages and unemployment benefits. In line with the macroeconomic theory, decreasing demand results in increasing unemployment. Real wage rigidity counteracts the contraction of demand and reduces the level of unemployment. The nonlinearity of the model is defined by a positive feedback between the negative growth in aggregate demand and unemployment. We have found all these driving forces irrelevant since the rate of unemployment is driven by the change in labor force. Our model is a linear (and likely lagged) one and any change in labor force is proportionally mapped into unemployment.

The reminder of the paper is organized in four sections. Section 1 formally introduces the model as obtained and statistically tested in previous studies (Kitov, 2006, 2007; Kitov and Kitov, 2010). In many countries, the US and Japan among others, the generalized link between labor force and two dependent variables can be split into two independent relationships, where inflation apparently does not depend on unemployment. In France, only the generalized model provided an adequate description of the evolution of both dependent variables since the 1960s. In Section 1, we also present and characterize various estimates of inflation, unemployment and labor force in Australia.

Section 2 discusses the Phillips curve in Australia and reveals the break in relevant time series around 1994. This year introduces a structural break in all relationships estimated in this study. In Section3, the linear link between labor force and unemployment is modeled using annual and monthly measurements of both variables. Instead of poorly constrained LSQ methods we apply a simplified version of the 1-D boundary element method – cumulative curves. Both empirical relationships are accurate.

Section 4 is devoted to the link between the rate of inflation and labor force. We also use the method of cumulative curves in order to estimate all coefficients in the relevant empirical relationship. Finally, Section 5 presents the generalized link between inflation, unemployment and labor force. The best fit model provides an accurate prediction of inflation as a function of labor force and unemployment.

1. **The model and data**

As originally defined by Kitov (2006), inflation and unemployment are linear and potentially lagged functions of the change rate of labor force:

$$\pi(t) = A_1 dLF(t-t_1)/LF(t-t_1) + A_2 \tag{1}$$
$$UE(t) = B_1 dLF(t-t_2)/LF(t-t_2) + B_2 \tag{2}$$

where $\pi(t)$ is the rate of price inflation at time $t$, as represented by some standard measure such as GDP deflator (DGDP) or CPI; $UE(t)$ is the rate of unemployment at time $t$, which can be also represented by various measures; *LF(t)* is the level of labor force at time *t*; $t_1$ and $t_2$ are the time lags between the inflation, unemployment, and labor force, respectively; $A_1$, $B_1$, $A_2$, and $B_2$ are country specific coefficients, which have to be determined empirically in calibration procedures. These coefficients may vary through time for a given country, as induced by numerous revisions to the definitions and measurement methodologies of the studied variables, i.e. by variations in measurement units.

Linear relationships (1) and (2) define inflation and unemployment separately. These variables are two indivisible manifestations or consequences of a unique process, however. The process is the growth in labor force which is accommodated in developed economies (we do not include developing and emergent economies in this analysis) through two channels. First channel is the increase in employment and corresponding change in personal income distribution (PID). All persons obtaining new paid jobs or their equivalents presumably change their incomes to some higher levels. There is an ultimate empirical fact, however, that PID in the USA does not change with time in relative terms, i.e. when normalized to the total population and total income. The increasing number of people at higher income levels, as related to the new paid jobs, leads to a certain disturbance in the PID. This over-concentration (or "over-pressure") of population in some income bins above its "neutral" long-term value must be compensated by such an extension in corresponding income scale, which returns the PID to its original density. Related stretching of the income scale is the core driving force of price inflation, i.e. the US economy needs exactly the amount of money, extra to that related to real GDP growth, to pull back the PID to its fixed shape. The mechanism responsible for the compensation and the income scale stretching, should have some positive relaxation time, which effectively separates in time the source of inflation, i.e. the labor force change, and the reaction, i.e. the inflation.

Second channel is related to those persons in the labor force who failed to obtain a new paid job. These people do not leave the labor force but join unemployment. Supposedly, they do not change the PID because they do not change their incomes. Therefore, total labor force change equals unemployment change plus employment change, the latter process expressed through lagged inflation. In the case of a "natural" behavior of the economic system, which is defined as a stable balance of socio-economic forces in the society, the partition of labor force growth between unemployment and inflation is retained through time and the linear relationships hold separately. There is always a possibility, however, to fix one of the two dependent variables. Central banks are definitely able to influence inflation rate by monetary means, i.e. to force money supply to change relative to its natural demand. To account for this effect one should to use a generalized relationship as represented by the sum of (1) and (2):

$$\pi(t)+UE(t)= A_1 dLF(t-t_1)/LF(t-t_1)+B_1 dLF(t-t_2)/LF(t-t_2)+A_2+B_2 \qquad (3)$$

Equation (3) balances the change in labor force to inflation and unemployment, the latter two variables potentially lagging by different times behind the labor force change. Effectively, when $t_1 \neq 0$ or/and $t_2 \neq 0$, one should not link inflation and unemployment for the same year.

One can rewrite (3) in a form similar to that of the Phillips curve, but without autoregressive terms:

$$\pi(t) = C_1 dLF(t-t_1)/LF(t-t_1)+C_2 UE(t+t_2-t_1)+C_3 \qquad (4)$$

where coefficients $C_1$, $C_2$, and $C_3$ should be better determined empirically despite they can be directly obtained from (3) by simple algebraic transformation. The rationale behind the superiority of the empirical estimation is the presence of high measurement noise in all original time series. In some places, (4) can provide a more effective destructive interference of such noise than does (3). Consequently, the coefficients providing the best fit for (3) and (4), whatever method is used, may be different.

The principal source of information is the OECD database (http://www.oecd.org/) which provides comprehensive data sets on labor force, unemployment, GDP deflator (DGDP), and CPI inflation. We also use the estimates reported by the U.S. Bureau of Labor Statistic (http://www.bls.gov) for corroboration of the data on CPI, unemployment and labor force. As a rule, readings associated with the same variable but obtained from different sources do not coincide. This is due to different approaches and definitions applied by corresponding agencies. This diversity of definitions is accompanied by a degree of uncertainty related to the methodology of measurements. This uncertainty cannot be directly estimated but certainly affects the reliability of empirical relationships.

In addition to all these data quality problems, there is no compatibility in definitions and measurement procedures over time. All data provided by all statistical agencies have to be checked for artificial breaks. For Australia, the OECD (2008) reports the following:

> **Series breaks**: A new questionnaire was introduced in 2001 and employment and unemployment series were re-estimated from 1986. From April 1986, employment data include unpaid family workers having worked less than 15 hours in a family business or on a farm. Previously, such persons who worked 1 to 14 hours or who had such a job but were not at work, were defined as either unemployed or not in the labor force, depending on whether they were actively looking for work.

Several central banks, including the Reserve Bank of Australia, shifted their monetary policy to inflation targeting around 1994. This introduced a real structural break in underlying time series and the generalized dependence between three macroeconomic variables under study is mandatory to use.

To begin with, we introduce the estimates of all variables. There are two time series for inflation, unemployment and the level of labor force. Figures 1 displays the evolution of two principal measures of price inflation – GDP deflator, *DGDP*, and consumer price index, *CPI*. Both variables were published by the OECD. As has been already discussed, we consider the DGDP as a better representative of price inflation in a given country since it includes all prices related to the economy. The overall consumer price index, *CPI*, is fully included in the DGDP and thus its behavior represents only a larger part of the economy. Since labor force and unemployment do characterize the entire economy it is methodically better to use DGDP for any quantitative modeling. Figure 1 shows that the overall difference between the CPI and DGDP is minor but there are short periods of very large discrepancy: from 1984 to 1987, from 1993 to 1995, and since 2003.

The rate of inflation fell to the level of 0.04 $y^{-1}$ in 1991 and has been oscillating around this level since. Between 1974 and 1990, inflation was almost everywhere above 6% per year. This behavior is similar to that in many developed countries and our concept well explained both the peak and the fall in the U.S. using only the change in labor force.

Figure 2 depicts two estimates of the rate of unemployment as reported by the OECD and the U.S. BLS. Surprisingly, the difference between these curves is almost negligible. There are two sharp peaks – in 1984 and 1994, i.e. the years with potential (artificial and real) structural breaks in all time series. The highest rate of unemployment in Australia was at the level of 11% in 1993, and the lowermost one was around 1% before 1965. This extremely low rate of unemployment was likely related to definition and measuring problems. We prefer to refrain from quantitative modeling of the period prior to 1970.

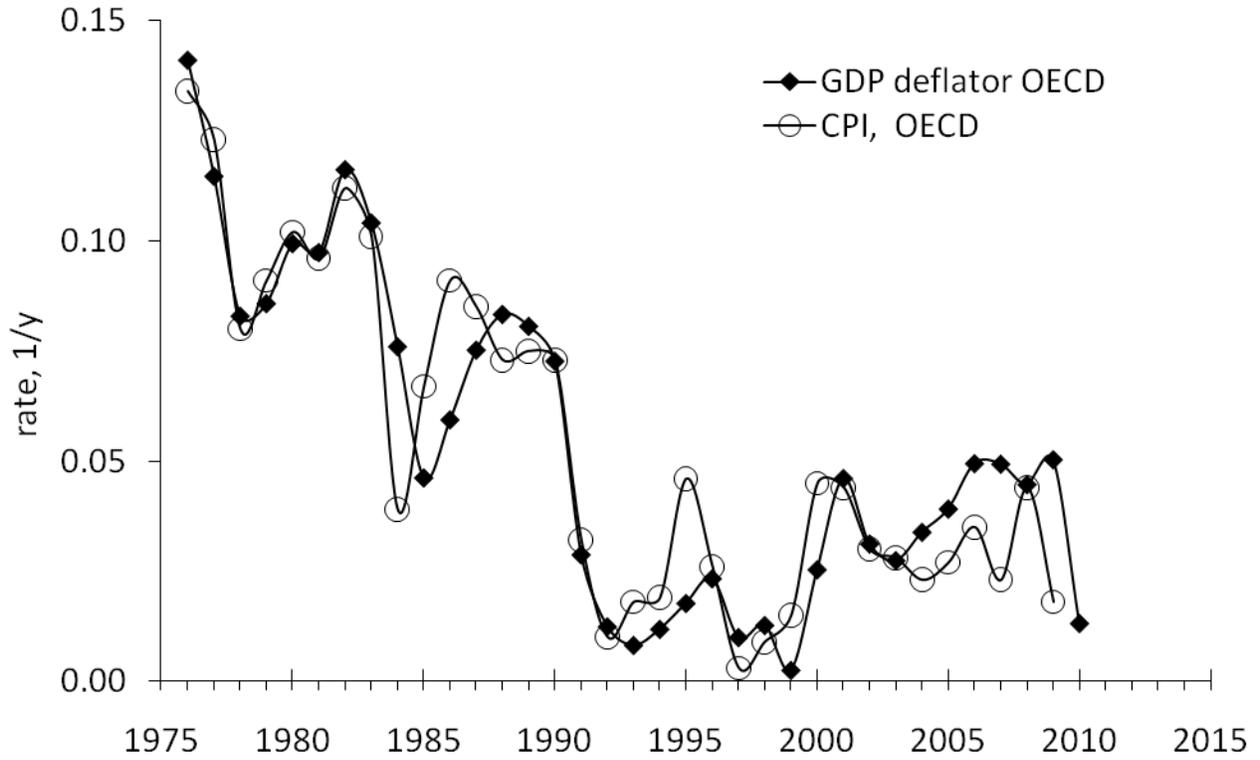
**Figure 1. Comparison of CPI inflation and GDP deflator in Australia, both reported by the OECD.**

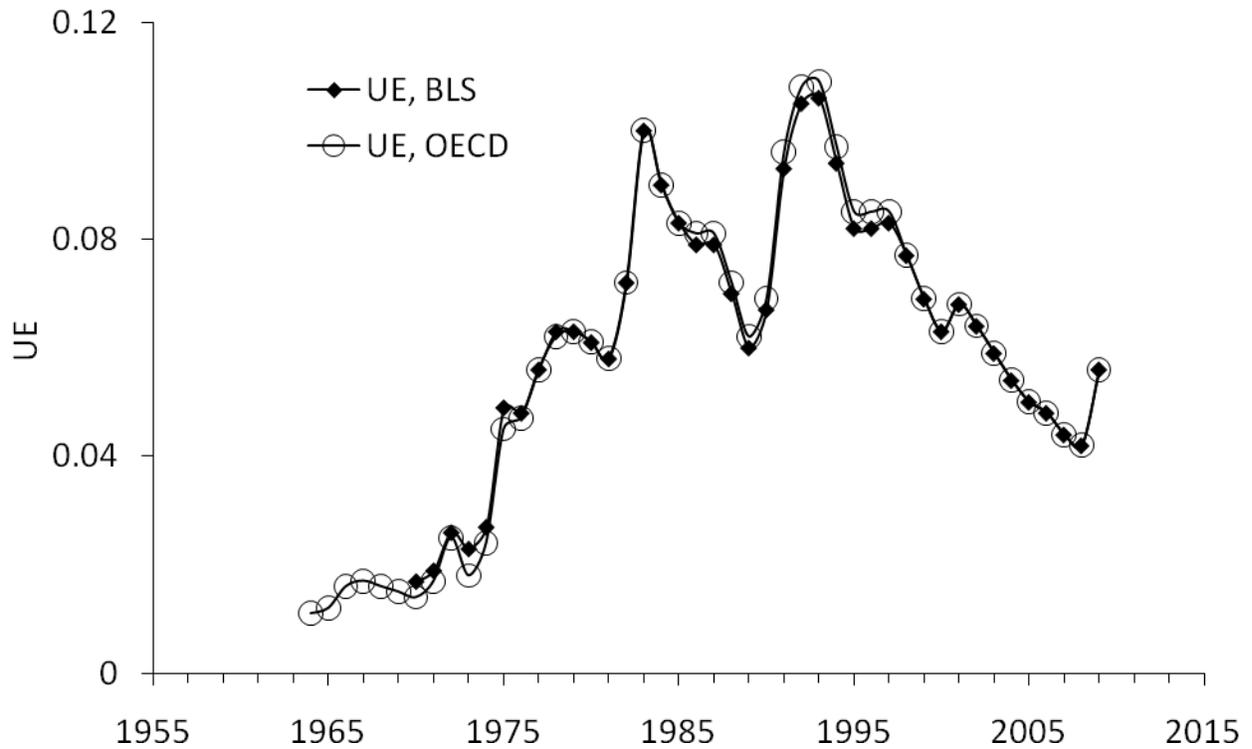
**Figure 2. Comparison of two estimates of unemployment according to the U.S. BLS and OECD definitions.**

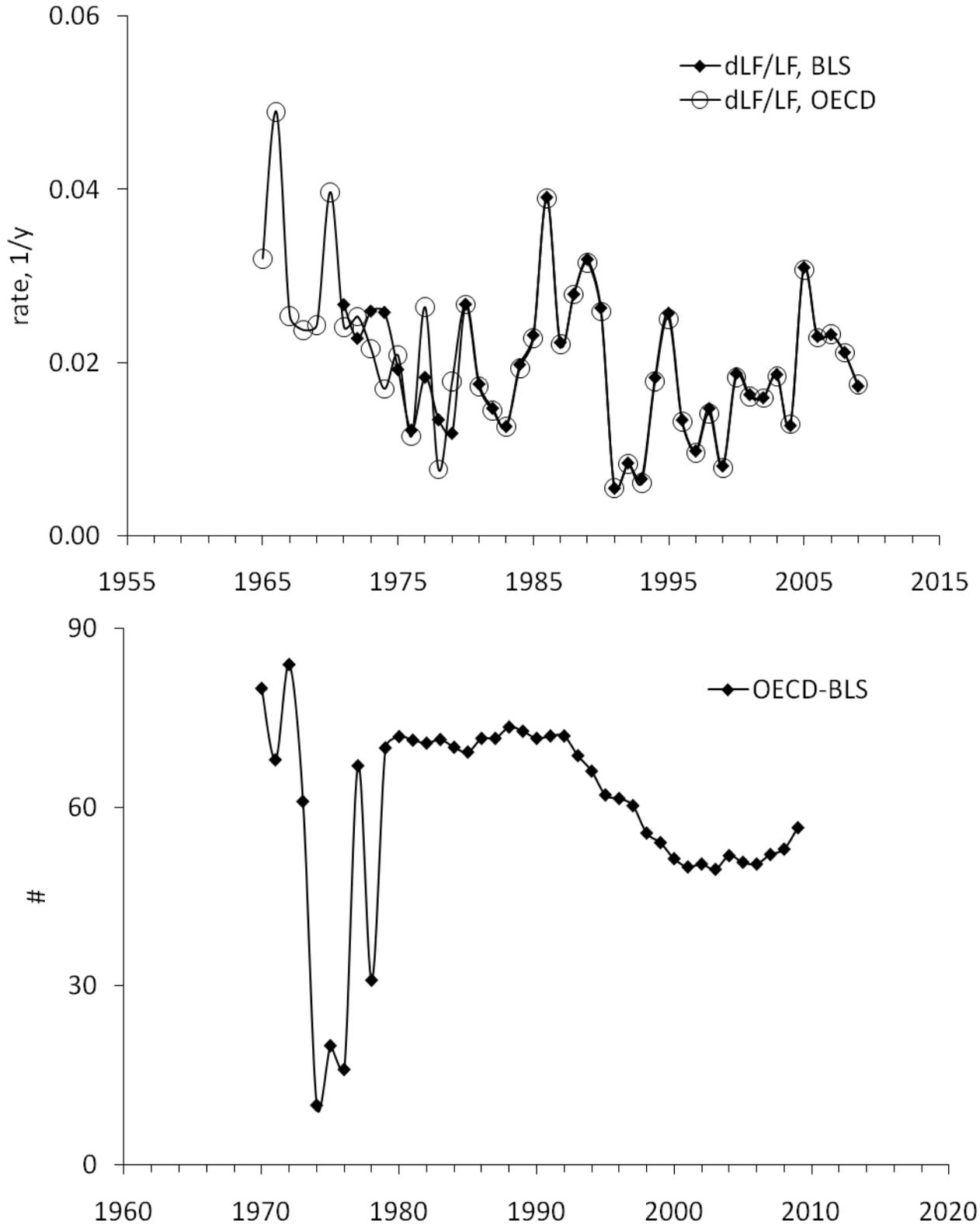

**Figure 3.** Comparison of two estimates of the change rate of labor force level – according to the OECD and U.S. definition (BLS). The difference in the lower panel shows a higher volatility before 1980.

The rate of change in labor force in |Figure 3 also has two representations: the OECD and BLS. Both time series are very close after 1980 with the overall difference between 45,000 to 60,000. All in all, this is an excellent agreement and the rate of change is practically identical. The period before 1980 is not so attractive for quantitative modeling – the difference between the BLS and OECD time series is very volatile. This observation may be related to the long-term iterative process of revision to labor force definition. Hence, one should not expect any good quantitative agreement between observed and predicted time series before 1980 – the level of labor force, and thus, the rate of change both have very high uncertainty.

The main task of this study is to estimate the best fit empirical relationships between these three variables. By definition, all relationships are linear and potentially lagged. Therefore, this task does not seem a difficult one. As a rule, we prefer to use cumulative curves instead of linear regression since the latter provides heavily biased estimates of the slope when both variables have high uncertainty.

## 2. The Phillips curve

In its simplest form, the Phillips curve is a statistical link between price inflation and unemployment. It can be represented as a scatter plot or as time series of the rate of unemployment in Australia and the rate of CPI inflation reduced to the unemployment time series by a linear relationship, as displayed in Figure 4. The period between 1974 and 1994 shows a good agreement with $R^2$=0.76. After 1995 the observed and predicted curves diverge. This is likely related to the new monetary policy of the Australian central bank associated with inflation targeting.

Figure 4 shows that the Phillips curve does not exist in Australia for the entire period between 1974 and 2009 as a single relationship between inflation and unemployment. As discussed above, we have limited quantitative modeling to the period of accurate measurements presented by the Australian Bureau of Statistics. The discrepancy between the observed and predicted curves in Figure 4 can manifest the absence of any relation between inflation and unemployment since 1995 or represent a structural break. The latter case can be reduced to a new linear relationship after 1994. Therefore, our next task is to estimate both coefficients of linear equation for the period after 1994.

Theoretically, linear regression is the technique to estimate the coeffcients. However, both variables, especially the CPI time series, are characterized by significant uncertainty. When both variables have uncertainty, the method of linear regression underestimates the slope. Because of this bias we prefer to estimate the coefficients by fitting cumulative curves of the measured variables. When a linear link between the studied variables does exist the cumulative curves must converge over time in realtive terms. In the absence of the link, no convergence is possible and the cumulative curves will give a clear and strong signal of the discrepancy. The superiority of cumulative values may be illustrated by a simple example. One can calculate average velocity of a car by integrating instant velocity estimates or dividing total path by travel time. The latter value is more accurate than the integral one and the relative accuracy only increases with travel time.

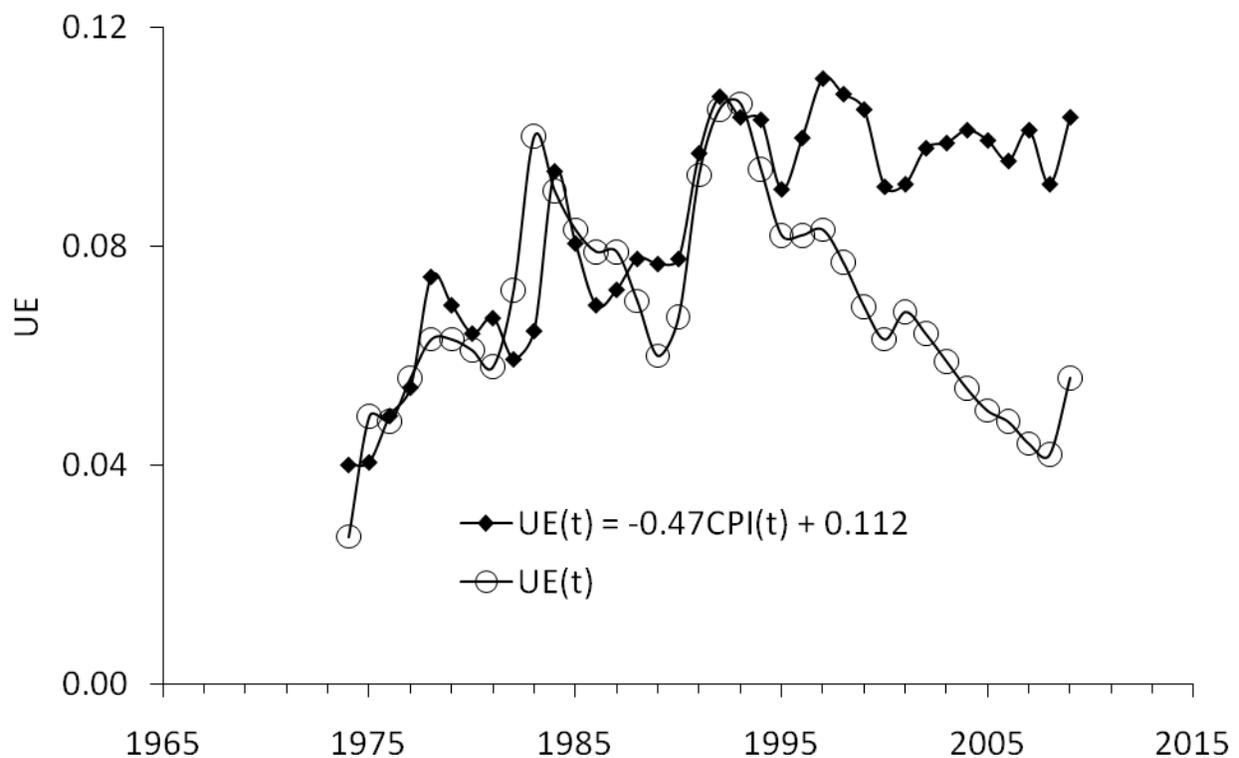

**Figure 4. Comparison of the measured unemployment (OECD) and that predicted from the CPI time series by relationship shown in the Figure. The curves are close between 1974 and 1994 with $R^2=0.76$ for this period. The following deviation might result from changes in monetary policy after 1994 and also be associated with revisions to corresponding definitions and measuring procedures.**

Figure 5 presents the results of the trial-and-error method applied to the observed and predicted cumulative curves. In the upper panel, the annual readings are plotted. The predcited curve is volatile, and thus, is smoothed by a three-year moving average, MA(3). As required, the cumulative curves in the lower panel are very close over the entire period. As before, we sought only visual fit between the curves and did not exercise any formal minimization. Our previous experience shows that formal statistical approach gives practically no improvement on visual fit. Finally, the Phillips curve for Australia has a structural break in 1994 and can be represented by two linear relationships before and after 1994:

$$UE(t) = -0.47CPI(t) + 0.112; \; t<1995$$
$$UE(t) = -1.5CPI(t) + 0.105; \; \; t>1994 \tag{5}$$

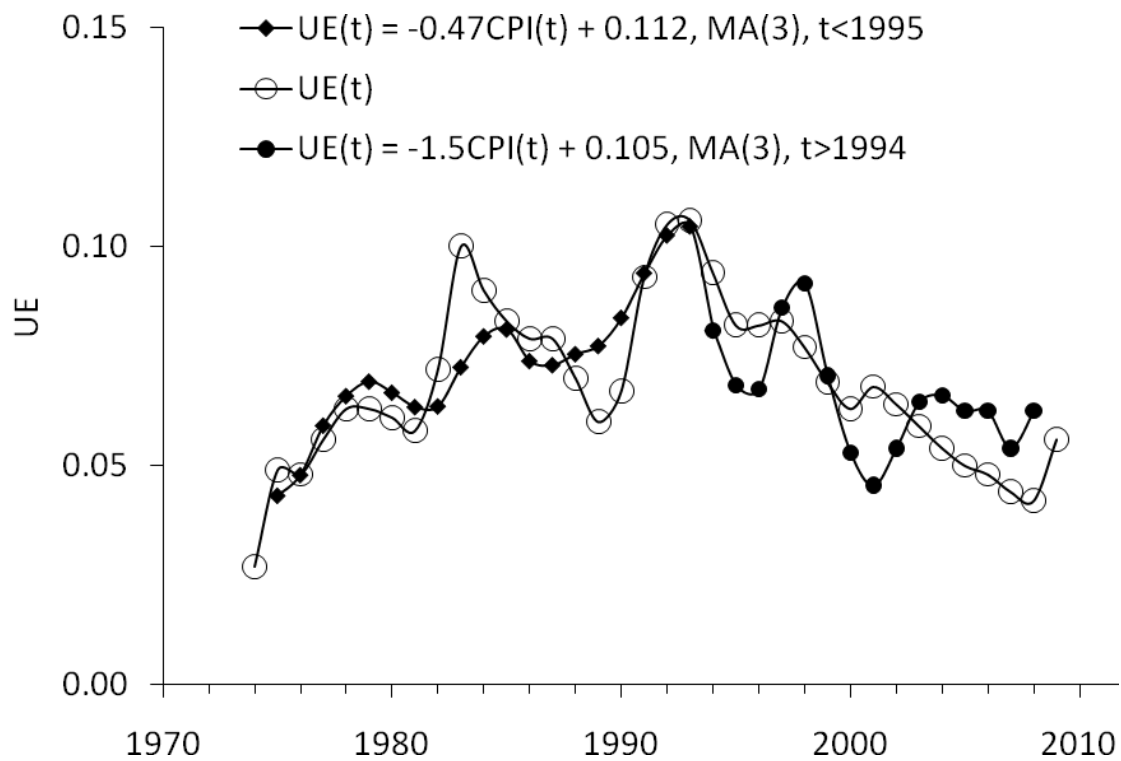

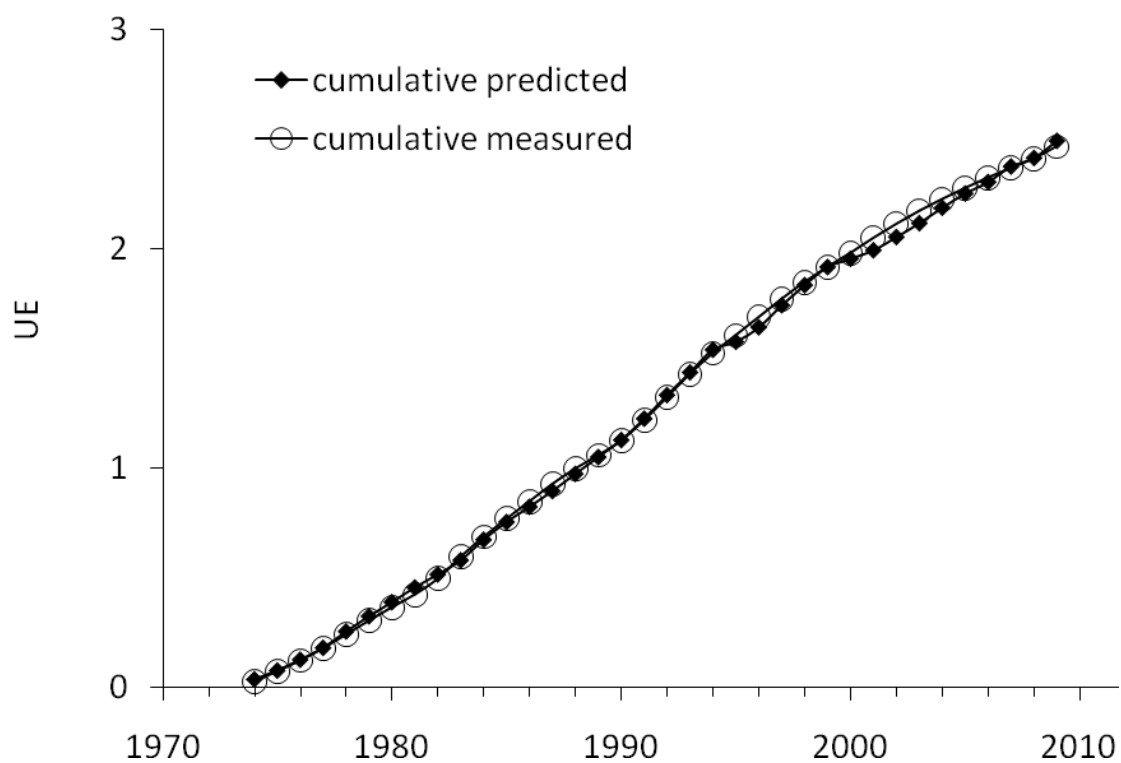

**Figure 5. The Phillips curve with a structural break in 1994. Coefficients in the defining relationships are obtained by trial-and-error methods from the cumulative curves in the lower panel.**

In any case, the existence of the canonical Phillips curve in Australia, even with a structural break is good news for our concept. The pair inflation/unemployment must demonstrate a piecewise linear relationship as a consequence of the generalized relationship.

### 3. Unemployment as a linear function of the change in labor force

From the Phillips curve it is only one step to the dependence of unemployment of the change in labor force. Actually, we replace the rate of inflation with the rate of labor force change in (5) and then have to estimate new coefficients: it has been empirically revealed and statistically tested that the rate of unemployment in developed countries is a linear (and lagged) function of the change in labor force.

As expected, the same relationship to be valid for Australia. The estimation method is as before – the trail-and-error one. For the annual readings in Figure 6, we do not use the cumulative curve approach and fit only peaks and troughs. The best-visual-fit equations for the period before and after 1994 are as follows:

$$UE(t) = -2.1dLF(t)/LF(t) + 0.13, \; t<1995$$
$$UE(t) = -2.1dLF(t)/LF(t) + 0.098, \; t>1994 \tag{6}$$

Because of high-amplitude oscillations in the original time series for the raet of labor force change, *dLF/LF*, we have to smooth it by MA(3). The slope in (6) is negative. Therefore, any increase in the level of labor force is reflected in a proportional and simultaneous fall in the rate of unemployment. This is a fortunate link – more jobs is equivalent to less unemployed. However, when the level of labor force does not change with time the rate of unemployment is very high – around 12%. Hence, Australia has to keep a higher rate of labor force growth in order to retain the rate of unemplyment at the current low level around 5%. Obviously, the jump from 4.5% in 2008 to 5.6% in 2009 was induced by the fall in labor force observed since 2007 (see Figure 3).

It is worth noting that (6) implies a nonlinear dependence on the rate of particiaption in labor force. For a given absolute change in the level of labor force in Australia, say 100,000 per year, the reaction of unemployment with be different for the rate of participation 60% and 65%. The higher is the participation rate the lower is the change rate, *dLF/LF*, and thus the change in the rate of unemployment. Actually, the participation rate in Australia has been increasing from ~62% in the 1970s and 1980s to 65.5% in the 2000s. It will be a difficult task to retain the rate of unemployment at the current low level – it is likely that the rate of participation is approaching the peak level and will start to decline in the near future.

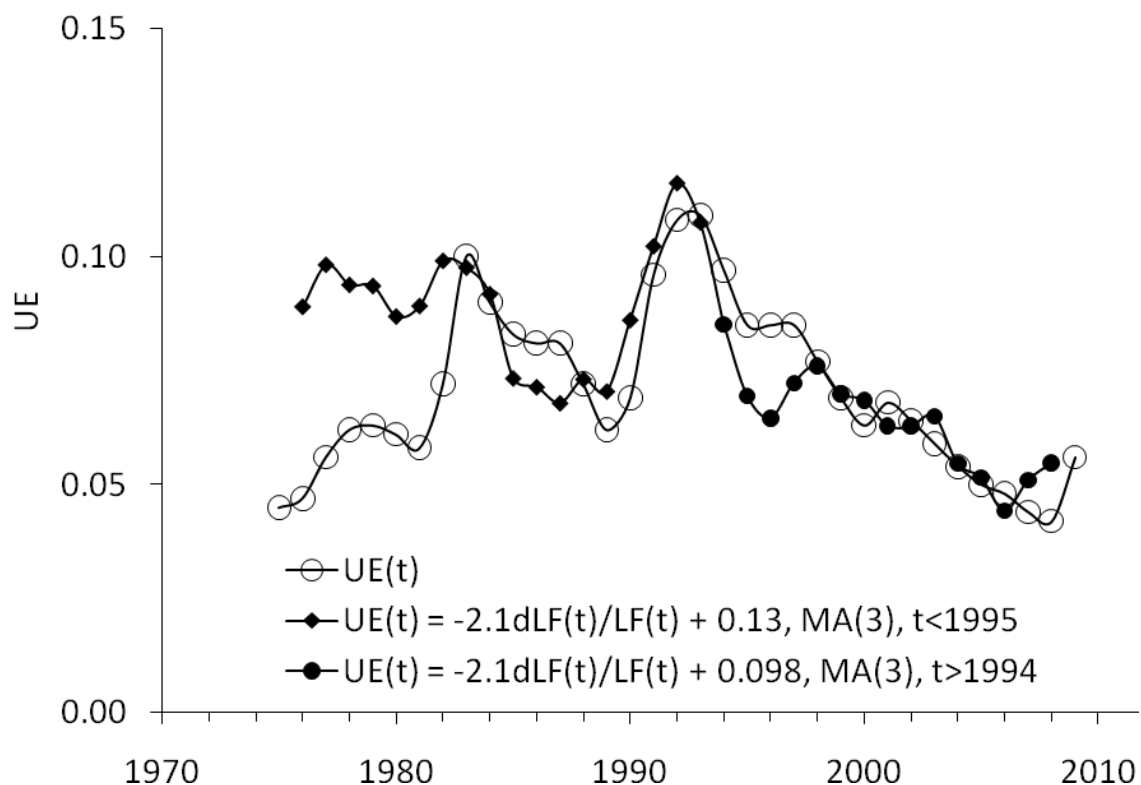

**Figure 6. Annual estimates of the rate of unemployment in Australia: measured vs. predicted from the change in labour force.**

Monthly readings of both variables are also available. Thus, one can apply the trial-and-error method to cumulative unemployment as published by the ABS at a monthly rate (see Figure 7). The best visual fit allows estimating all coefficients:

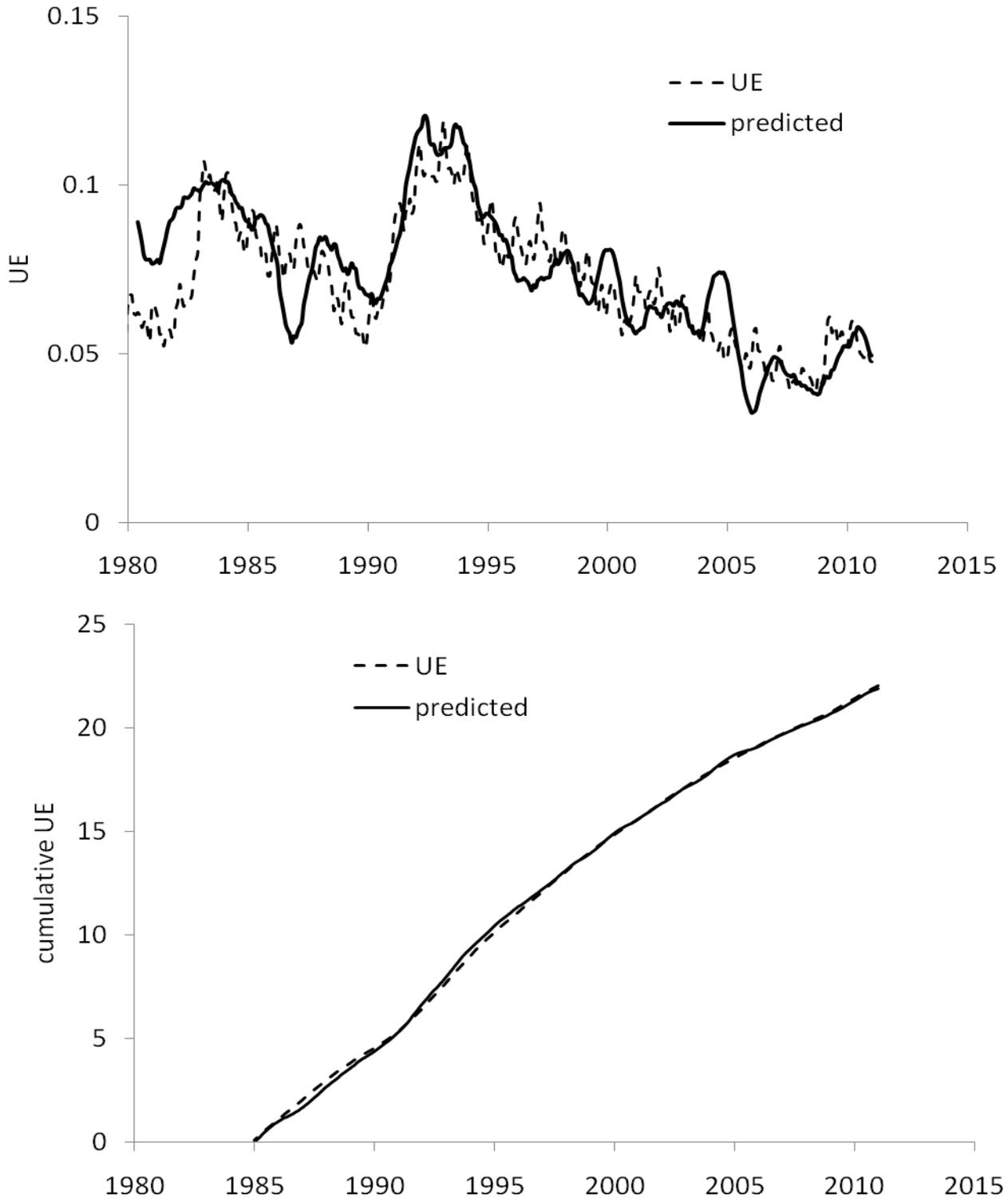

**Figure 7.** *Upper panel*. Monthly estimates of the rate of unemployment in Australia and that obtained from labor force using (7). Due to high-amplitude fluctuations in the monthly estimates of *dLF/LF*, the predicted curve is smoothed by a twelve-month moving average, MA(12). *Lower panel*. Cumulative values of the observed and predicted curves in the upper panel. Notice the excellent agreement between the cumulative curves.

$$UE(t) = -1.77dLF(t)/LF(t) + 0.124; \quad t<1995$$
$$UE(t) = -2.1dLF(t)/LF(t) + 0.0977; \quad t>1994 \qquad (7)$$

Again, because of the change in monetary policy around 1995 we have to split the modeled period into two segments: before and after 1994. The slope in the linear relationship is the same over the entire period. This model is somewhat different from (6), which is based on annual readings. This is likely associated with the quick-and-dirty approach used to estimate (6). The method of cumulative curves gives much better estimates.

Figure 3 illustrates the limits of accurate quantitative modeling. Before 1980, one should not expect any accurate predictions. In Figure 7, we seek for the best visual fit between cumulative curves after 1985. After 1995, our model has a superior predictive power. As a compromise, we carried out several statistical tests of the observed and predicted time series as well as the model residual for the period since January 1987. There are 288 readings used.

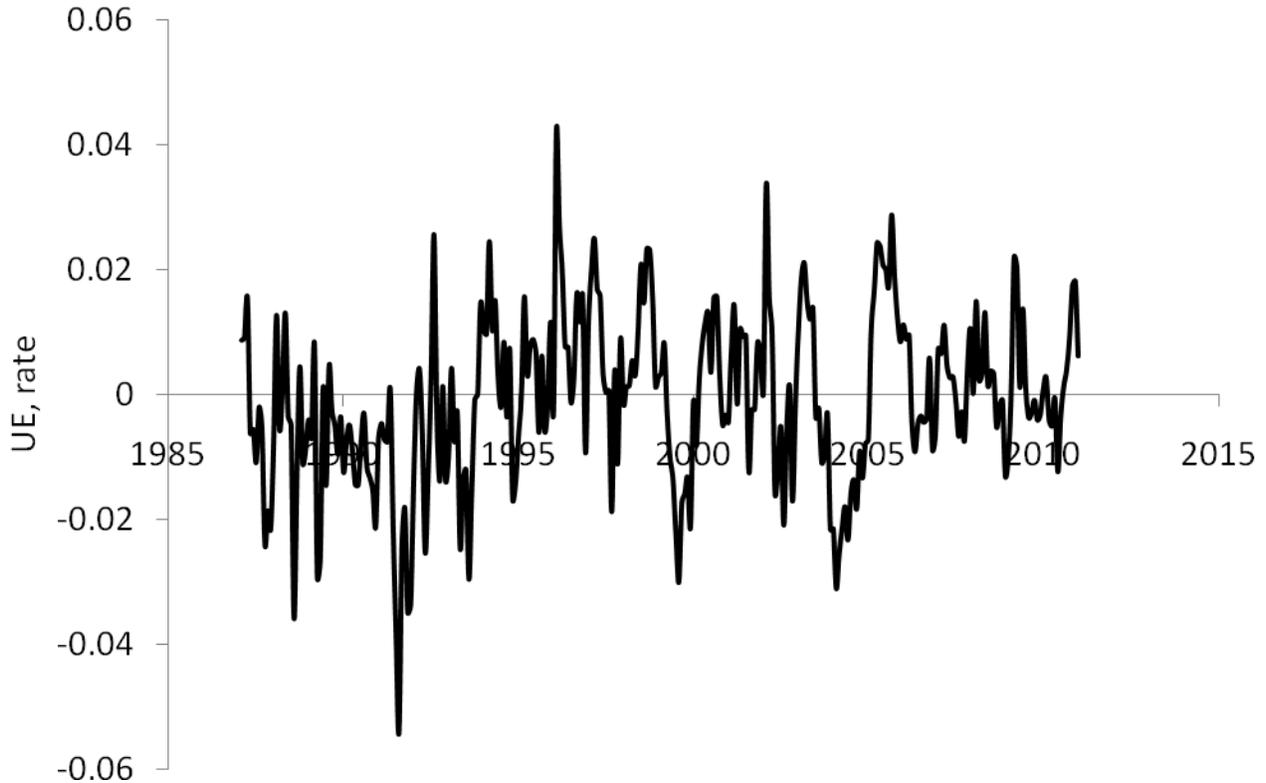

**Figure 8. The residual model error for the monthly unemployment estimates.**

At first, the residual model error shown in Figure 8 was tested for unit roots. The augmented Dickey-Fuller test gave z(t)=-7.56 with the 1% critical value of -3.46. The Phillips-Perron test for unit roots gave z(ρ)=-104 and z(t)=-7.88, with the 1% critical value of -20.3 and -3.46, respectively. Thus, one can reject the null hypothesis of a unit root in the residual time series. This confirms that the predicted time series is cointegrated with the observed one, as the Engle-Granger test for cointegration requires. The Johansen test for cointegration applied to the observed and predicted time series gave a robust rank 1. Because the cumulative approach provides the overall detrending of the model residual we selected trend specification "none". Econometrically, there exists a cointegrating relation between the rate of unemployment and the change in labor force in Australia. This makes the result of linear regression of the monthly reading an unbiased one: $R^2$=0.81; for cumulative curves, $R^2$=0.999.

All in all, the agreement between the annual and cumulative curves is excellent. We have smoothed the predicted curve by a twelve-month moving average is order to suppress high-amplitude fluctuations in the monthly estimates of dLF/LF. One can predict the rate of unemployment at any time horizon using labor force projections. We have failed to find any projection published by the Australian Bureau of Statistics except the one between 1999 and 2016. Unfortunately, this projection was all wrong and heavily underestimated the growth in labor force. It predicted the level of labor force in 2016 at 10,800,000. In December 2010, the level of labor force was 12,132,900. This is good news, however. According to (7), a higher rate of labor force growth results in a lower rate of unemployment.

## 4. Inflation as a linear function of the change in labor force

The existence of a deterministic link between labor force and price inflation has been proved for many countries. Here we are following the same estimation procedure as for unemployment above and other developed countries. We start with the annual readings of GDP deflator, *DGDP,* reported by the OECD. According to the change in definition of labor force in 1986, which first affected the estimates for 1985, we have divided the period after 1978 (the start of reliable measurements as reported by the Australian Bureaus of Statistics) into two segments and obtained the following empirical models:

$$DGDP(t) = 4.2dLF(t)/LF(t) – 0.042 \quad t>1984$$
$$DGDP(t) = 7.8dLF(t)/LF(t) – 0.024 \quad t<1985 \qquad (8)$$

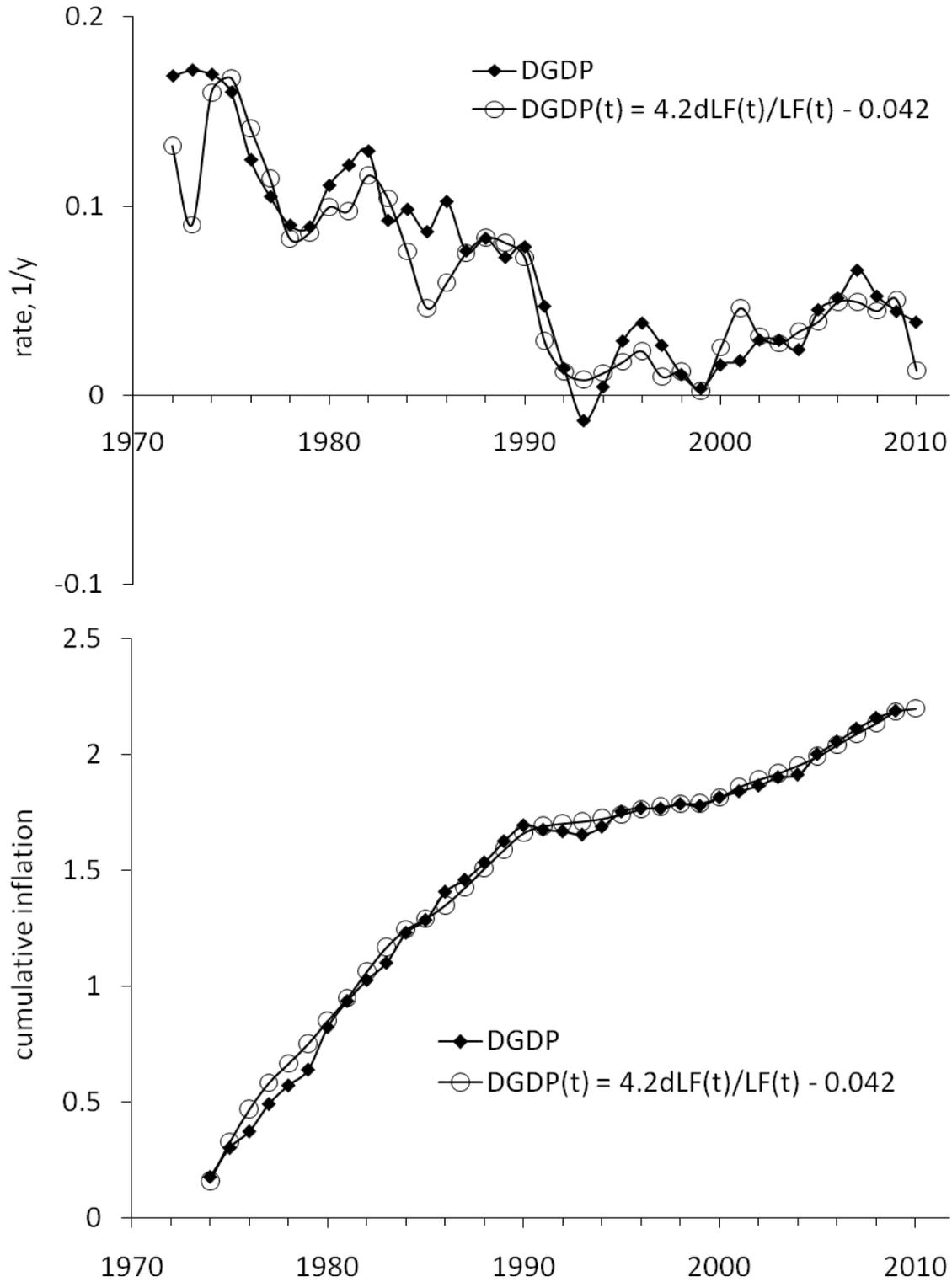

**Figure 9. Modeling the cumulative GDP deflator as a function of the change rate of labor force level. The break in 1985 is explained by the changes in definition the labor force definition and corresponding measurement procedure.**

Figure 9 displays the observed DGDP curve and that predicted according to (8). Both, dynamic and cumulative curves are in agreement. For the period between 1974 and 2009, the goodness of fit is very high: $R^2_{dyn}$=0.62 and $R^2_{cum}$= 0.996, respectively. If to consider that (8) does not use any autoregressive properties of inflation, which usually bring between 80% and 90% of the explanatory power in the mainstream models, this link is almost a deterministic one. Later in this Section we show that the Australian times series of inflation and the change in labor force are cointegrated.

Figure 10 illustrates the benefits of cumulative approach. The absolute and relative errors decrease with time. Despite the annual levels of price and labor force are not measured more accurately with time the overall change in the level is measured better and better. As a consequence, the observed and predicted cumulative curves, i.e. the overall changes in price and labor force, do converge. They become indistinguishable, i.e. there exists a deterministic link between them.

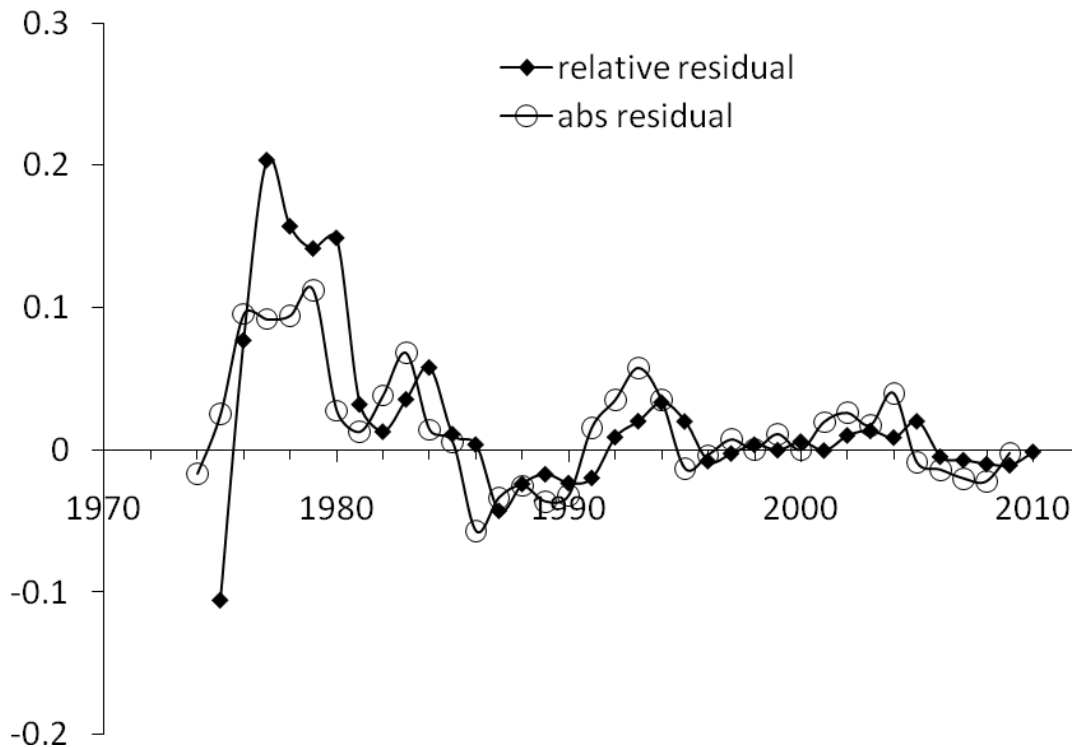

**Figure 10. Absolute and relative modeling error for the cumulative inflation in Figure 5. The curves converge in relative terms and one can replace the price deflator with the growth in labor force with the accuracy incrasing with time.**

The annual time series are relatively short with only 36 readings between 1974 and 2009, as shown in Figure 9. It is well known that small samples cannot provide robust statistical estimates and inferences. Fortunately, the OECD also reports quarterly estimates of inflation and labor force. As a rule, monthly and quarterly data are noisy because of measurement errors. For the Australian time series the overall measurement accuracy is not poor and we have obtained the estimates of coefficients in the linear link between the change rate of the GDP deflator, *DGDP* (annualized Q/Q), and *dLF/LF*:

$$DGDP(t) = 3.3dLF(t)/LF(t) - 0.026 \quad t>1984$$
$$DGDP(t) = 6.5dLF(t)/LF(t) - 0.021 \quad t<1985 \tag{9}$$

Figure 11 presents both the quarterly and cumulative curves of observed and predicted inflation. The agreement between the cumulative curves is excellent, but both time series are noisy. Obviously, the quarterly estimates of labor force and GDP deflator, in Australia and other developed countries, are not accurate and the total change from quarter to quarter may be less than the measurement uncertainty. The general resemblance is good, however. Peaks and troughs are well described but their amplitudes oscillate fast.

The cumulative inflation, i.e. the progressive sum of the (annualized) quarterly inflation values, has reached the level of 5.5, with the highest quarterly value of 0.12. Hence, the quarterly change in the cumulative value is less than 0.5%. The largest deviation between the cumulative curves is observed in 2010, when the measured DGDP dropped below zero line. This sharp deviation is based on preliminary estimates for 2010 and is subject to further revisions.

All in all, the current deviation does not change the final result of statistical tests for cointegration. According to the first step in Engle-Granger test for cointegration, the residual error of linear regression should not have unit roots. Figure 12 depicts the model residual as obtained using the method of cumulative curves (see Figure 11), which we consider as an equivalent of the regression residual error. For 122 readings between the second quarter of 1980 and the third quarter of 2010, the augmented Dickey-Fuller (DF) test gave z(t)=-4.40 with the 1% critical value of -3.50. The DF-GLS test rejects (1% critical value) the null of a unit root for all lags from 1 to 12 (quarters) except lag=4. The Phillips-Perron test for unit roots gave z(ρ)=-36.8 and z(t)=-4.54, with the 1% critical value of -19.87 and -3.50, respectively.

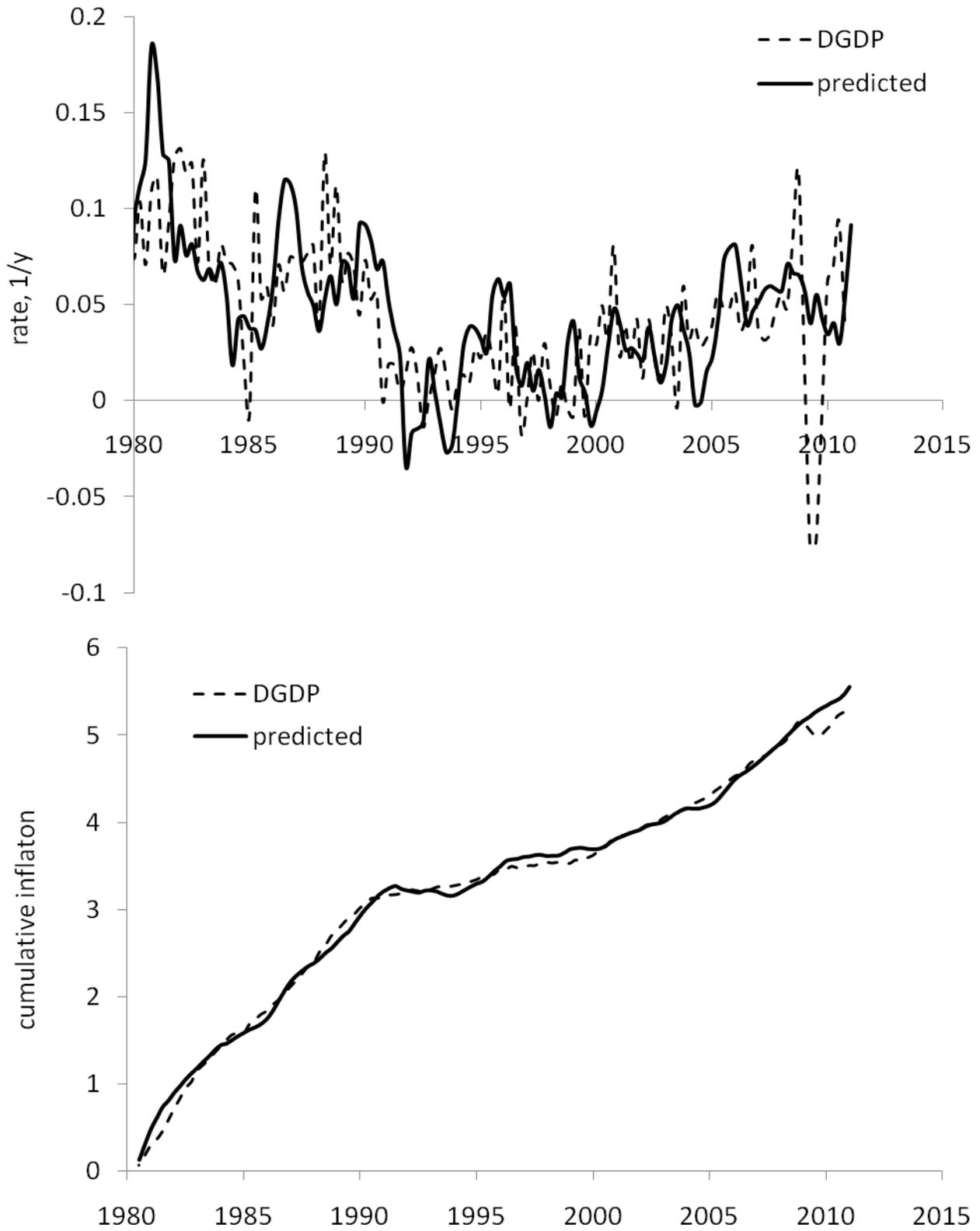

**Figure 11. Modeling the quarterly and cumulative DGDP estimates between the first quarter of 1980 and the third quarted of 2010.**

Therefore, the tests for unit roots prove that the predicted time series is cointegrated with the observed one. The Johansen test for cointegration checks the rank of the relation between the involved variables and thus the presence of cointegration relations. For two variables, rank one is equivalent to the existence of one cointegrating relation between these variables. When applied to the observed and predicted quarterly time series of inflation in Australia the Johansen test gave a robust rank 1 (trend specification "none"). Econometrically, there exists a long term equilibrium relation between the rate of unemployment and the change in labor force in Australia with a break in 1994. This makes the result of linear regression of the quarterly readings an unbiased one: $R^2=0.48$. For cumulative curves, $R^2=0.99$.

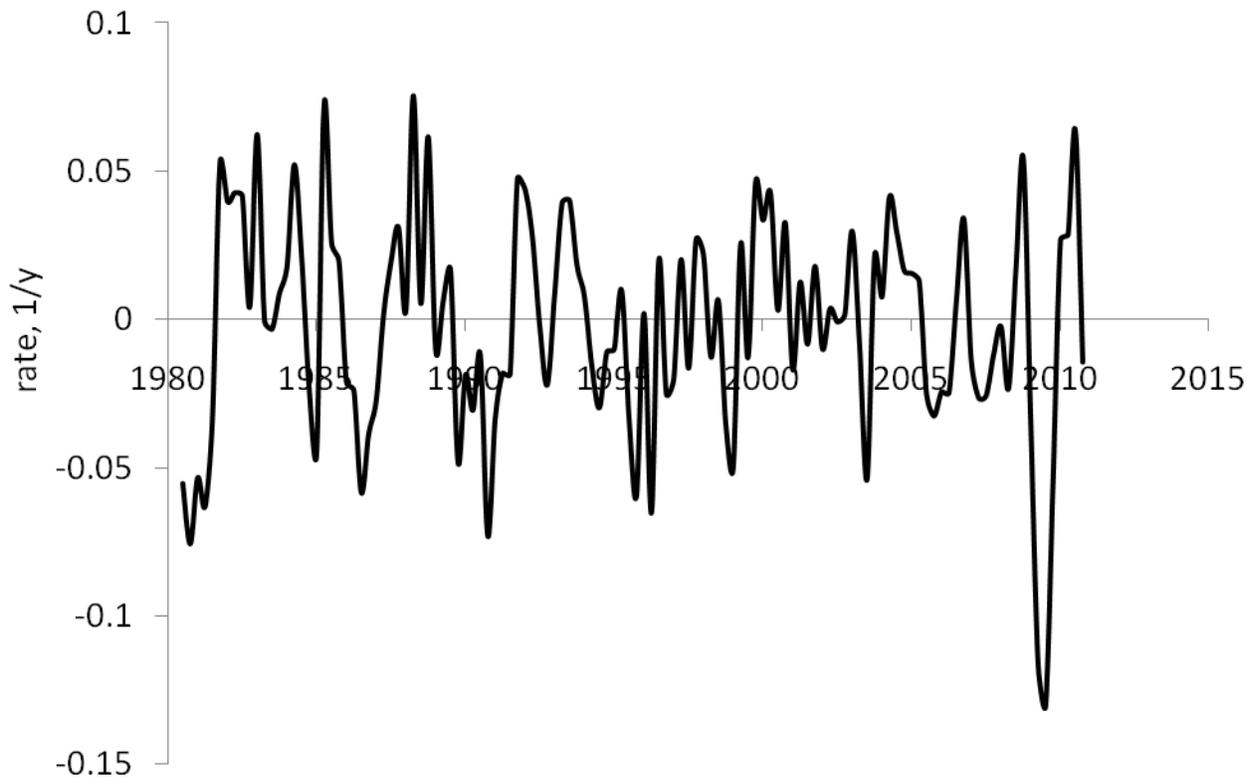

**Figure 12. The residual model error for the quarterly DGDP estimates.**

## 5. The generalized model

We have estimated several individual links between labor force, unemployment and inflation. Both relations to labor force are cointegrated, as the Engle-Granger and Johansen tests have shown. In this situation, the generalized model is a straightforward one and somewhat obsolete. However, we have estimated this model as well for methodological purposes and for the completeness of our concept. Because of the aforementioned breaks in 1995 and the break in 1984 we split the entire modeling period into three segments:

$$CPI(t) = 3.9dLF(t)/LF(t) + 0.88UE(t) - 0.1; \quad t>1995$$
$$CPI(t) = 3.9dLF(t)/LF(t) + 0.97UE(t) - 0.1; \quad 1984<t<1996$$
$$CPI(t) = 8.3dLF(t)/LF(t) + 0.97UE(t) - 0.1; \quad t<1985 \qquad (10)$$

Figure 13 presents the measured and predicted CPI inflation. As discussed above, the CPI does not represent the economy as a whole, and thus, the CPI evolution may be not a one-to-one reaction to the change in labor force. Figure 1 explains the difference - the CPI curve is more volatile and deviates from the DGDP one in many places. Having a good prediction of the GDPD, one should not expect the same accuracy of prediction with the CPI time series. This fact is clearly illustrated in Figure 13, where the predicted time series heavily deviates from the observed CPI inflation through the entire period. At the same time the rate of CPI inflation is a one-to-one function of the change in labor force and unemployment. Both coefficients are positive in (10), but we have already found that these variables evolve in opposite directions, as described by (6).

**Conclusion**

The rate of price inflation and unemployment in Australia is a one-off function of the change in labor force. This conclusion validates earlier models for many developed countries: the US, Japan, Germany, France, Italy, Canada, the Netherlands, Sweden, Austria, and Switzerland. Australia is one of few countries we did not study so far. The excellence of the obtained statistical and conceptual results compensates the delay in analysis.

Overall, we have established that there exist long term equilibrium relations the rate of labor force change and the rate of inflation/unemployment. The level of statistical significance of these cointegrating relations allows us to consider these links as deterministic ones, as adopted in physics. Unlike the mainstream models, no relation uses autoregressive properties of any macroeconomic variable under consideration. This does not make the rate of unemployment and inflation non-stochastic time series. The change in labor force includes a strong demographics component and thus is stochastic to the extent the evolution of population in a given country is stochastic. Since the level of labor force is a measurable value one does not need to estimate its stochastic properties – they are obtained automatically with routine measurements.

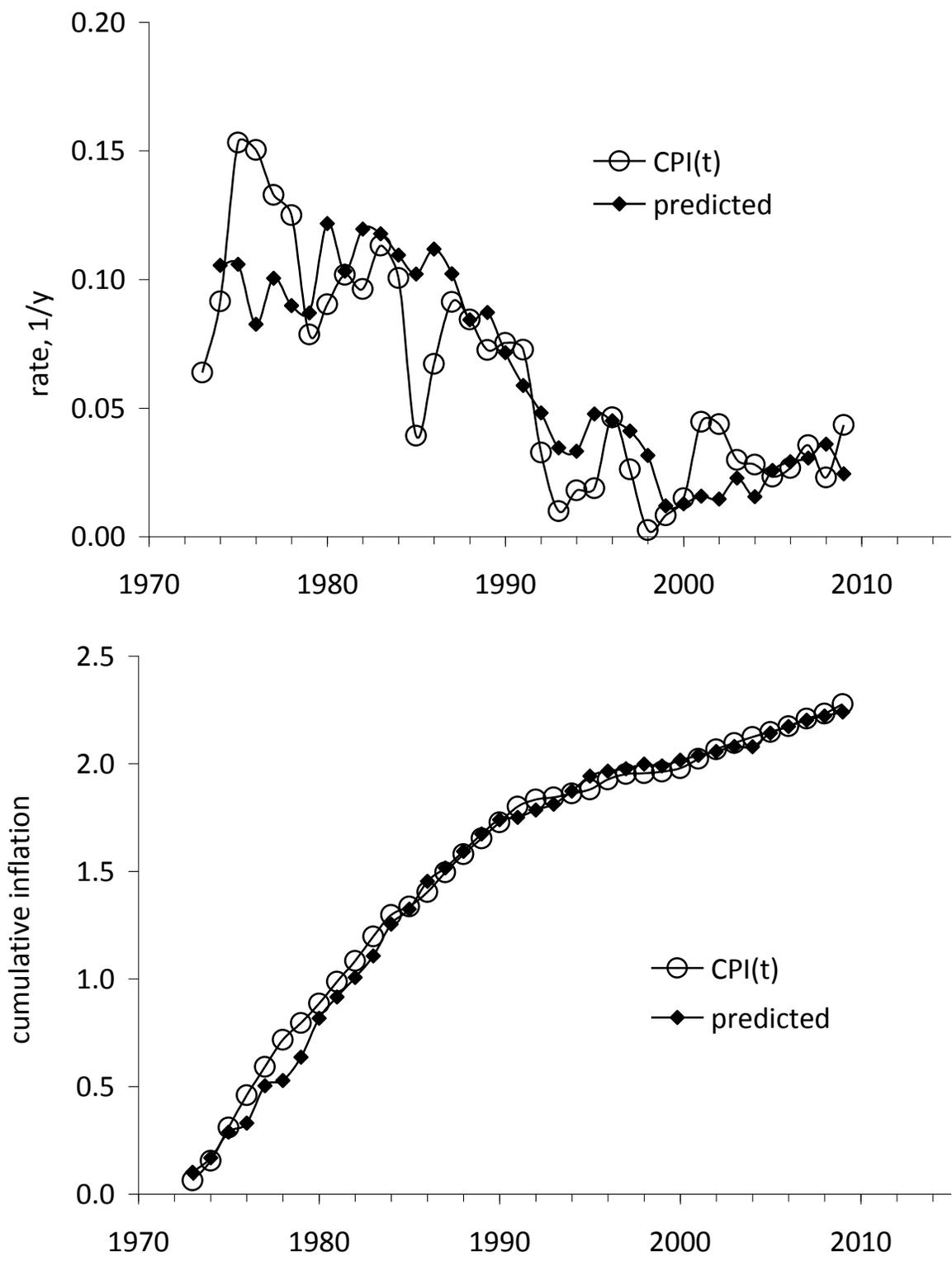

**Figure 13.** *Upper panel*: Illustration of the generalized relation between inflation, unemployment and the change rate of labor force leveling Australia. The CPI inflation is modeled using the change rate of labor force level and unemployment. *Lower panel*: Cumulative curves use to estimate all coefficients in defining relationships (10).

Unintentional but helpful outcome from the above modeling is the existence of the canonical Phillips curve in Australia, even with a structural break in 1995. The pair inflation/unemployment must demonstrate a piecewise linear relationship as a consequence of the generalized relationship we obtained in Section 5.